\documentclass[12pt]{article}
\usepackage{amssymb}
\usepackage{amsmath}
\usepackage{amsfonts}
\usepackage{epsfig}
\usepackage{anysize}
\marginsize{2cm}{2cm}{2.5cm}{3cm}
\date{}

\usepackage{hyperref}

\begin{document}
\title{\LARGE\bf Pascal (Yang Hui) triangles and power laws in the logistic map}

\author{Carlos Velarde\textsuperscript{1}, Alberto Robledo\textsuperscript{2}\\
\footnotesize 1. Instituto de Investigaciones en Matem\'aticas Aplicadas y en Sistemas, Universidad Nacional Aut\'onoma de M\'exico\\
\footnotesize 2. Instituto de F\'{i}sica y Centro de Ciencias de la Complejidad,
                  Universidad Nacional Aut\'onoma de M\'exico,\\ 
\footnotesize Apartado Postal 20-364, M\'exico 01000 DF, Mexico.
}

\maketitle

\abstract{%
 We point out the joint occurrence of Pascal triangle
 patterns and power-law scaling in the standard logistic map, or more
 generally, in unimodal maps. It is known that these features are present
 in its two types of bifurcation cascades: period and chaotic-band
 doubling of attractors. Approximate Pascal triangles are exhibited by
 the sets of lengths of supercycle diameters and by the sets of widths of opening
 bands. Additionally, power-law scaling manifests along periodic attractor
 supercycle positions and chaotic band splitting points. Consequently,
 the attractor at the mutual accumulation point of the doubling
 cascades, the onset of chaos, displays both Gaussian and power-law
 distributions. Their combined existence implies both ordinary and
 exceptional statistical-mechanical descriptions of dynamical
 properties.}

%

\section{Introduction}
The logistic map has played a prominent role in the development of the
field of nonlinear dynamics \cite{Schuster1}-\cite{Schroeder1}. The simplicity of
its quadratic expression and the richness and intricacy of the
properties that stem from it have captivated a large number of
scholars and students over decades. It has served as a standard source
for the illustration of nonlinear concepts such as: bifurcations,
stable and unstable periodic orbits, periodic windows, ergodic and
mixing behaviors, chaotic orbits and universality in the sense of the
Renormalization Group (RG) method \cite{Schuster1}-\cite{Schroeder1}. It has also
become a suitable model system for the exploration of
statistical-mechanical structures \cite{Robledo1}. All of these properties
are shared by one-dimensional unimodal maps where the quadratic
maximum is replaced by an extremum of general nonlinearity $z>1$
\cite{Beck1,Capel1}.  Here we concisely draw attention to the presence of
geometrical and scaling laws in the family of attractors generated by
the logistic map and to the consequences that these laws have in the
dynamical properties of their most interesting object of study: the
period-doubling transition to chaos. One-dimensional nonlinear maps,
like the logistic, are necessarily dissipative \cite{Schuster1} and they
settle after much iteration into attractors that may consist of a
finite set of points regularly visited or an infinite number of points
that may be irregularly (or chaotically) visited
\cite{Schuster1}-\cite{Schroeder1}.  The logistic map possesses an infinite family
of attractors that are connected via cascades of bifurcations at which
periods or number of chaotic bands duplicate \cite{Schuster1}-\cite{Schroeder1}.
These families of attractors display power laws associated with
attractor positions and Pascal Triangles associated with distances
between these positions.

  We recall that a Pascal Triangle is a
triangular arrangement of the binomial coefficients and that it
contains many remarkable numerical relations \cite{Pascal1}. Blaise Pascal
studied this array in the 17th century, although it had been described
centuries earlier by the Chinese mathematician Yang Hui, and then by
other Indian and Persian scholars. It is therefore known as the Yang
Hui triangle in China \cite{Yanghui1}. This triangle serves as the basis of
the De Moivre-Laplace theorem \cite{Moivre1} an earlier limited version of
the central limit theorem leading to the Gaussian distribution.

  We provide below a description of how these
properties arise in the logistic map and discuss their implications
for the mathematical structures in the dynamics of this nonlinear
system, itself a convenient numerical laboratory for the study of
statistical-mechanical theories \cite{Robledo1}.

\section{The bifurcation cascades of the logistic map}

  We briefly recall the
basic definitions of the superstable periodic attractors (or
supercycles) and the chaotic band-merging attractors (or Misiurewicz
points) \cite{Schuster1}-\cite{Schroeder1}. These have become convenient families
of attractors in formal descriptions of the bifurcation cascades of
unimodal maps, often illustrated by the logistic map
 $f_{\mu}(x)=1-\mu x^{2}$, $-1\leq x\leq 1$, $0\leq \mu \leq 2$.
 
The superstable orbits of periods $2^{n}$, $n=1,2,3, \ldots$, are located
along the bifurcation forks, i.e. the control parameter value 
$\mu=\bar{\mu}_{n}<\mu_{\infty}$ for the superstable
$2^{n}$-attractor is that for which the orbit of period $2^{n}$
contains the point $x=0$, where $\mu_{\infty }=1.401155189\ldots$ is
the value of $\mu$ at the main period-doubling accumulation
point. The positions (or phases) of the $2^{n}$-attractor are given
by $x_{m}=f_{\bar{\mu}_{n}}^{(m)}(0)$, $m=0,1, \ldots,2^{n}-1$.
The diameters $d_{n,m}$ are defined as
 $d_{n,m} \equiv x_{m} - f_{\bar{\mu }_{n}}^{(2^{n-1})}(x_{m})$ \cite{Schuster1}. See Fig.\ 1.
 Notice that infinitely many other sequences of superstable attractors appear
at the period-doubling cascades within the windows of periodic
attractors for values of $\mu > \mu_{\infty }$ \cite{Schuster1}.

\begin{figure*}[t] 
\centering
\epsfig{file=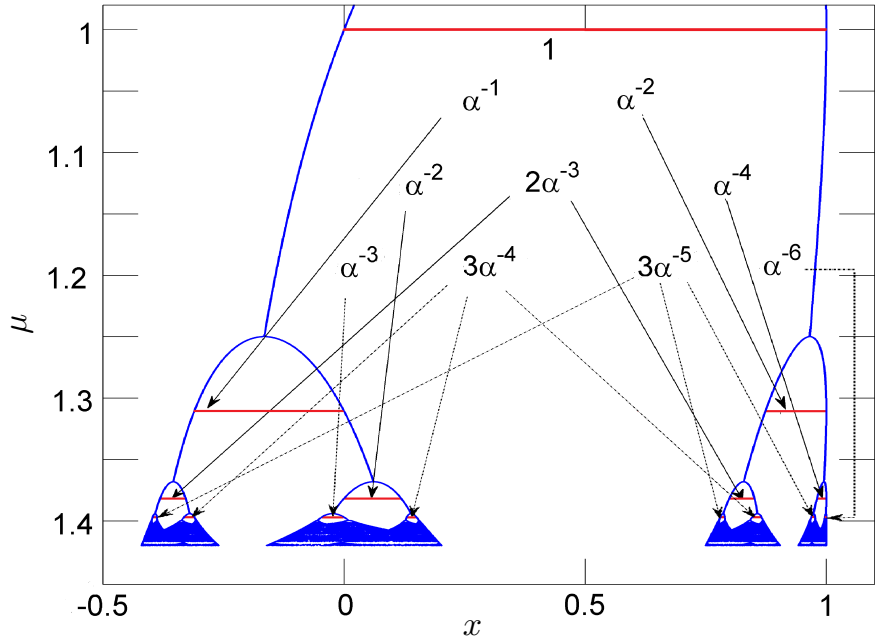, width=.67\textwidth} 
\caption{\scriptsize
   Sector of the period-doubling bifurcation
   tree for the logistic map $f_\mu(x)$ that shows the formation of a Pascal
   Triangle of diameter lengths according to the binomial approximation
   explained in the text, where $\alpha \simeq 2.50291$ is the absolute value of
   Feigenbaum's universal constant.}
\label{periodDoubling}
\end{figure*}

When $\mu $ is shifted to values larger than $\mu_{\infty }$,
 $\Delta \mu \equiv \mu -\mu_{\infty }>0$, the attractors are chaotic and
consist of $2^{n}$ bands, $n=0,1,2,\ldots$, where
 $2^{n}\sim \Delta \mu^{-\kappa }$, $\kappa =\ln 2/\ln \delta $, and
 $\delta =4.669201609102\ldots$ is the universal constant that measures both the rate of
convergence of the values of $\mu =\bar{\mu}_{n}$ to $\mu_{\infty }$ at
period doubling or at band splitting points or Misiurewicz points
\cite{Beck1}. The Misiurewicz ($M_{n}$) points are attractor merging
crises, where multiple pieces of an attractor merge together at the
position of an unstable periodic orbit. The $M_{n}$ points can be
determined by evaluation of the trajectories with initial condition
$x_{0}=0$ for different values of $\mu$, as these orbits follow the
edges of the chaotic bands until at $\mu =\hat{\mu}_{n}$,
the control parameter value for the $M_n$ point,
 the unstable orbit
of period $2^{n}$ reaches the merging crises \cite{Grebogi1}. See Fig.\ 2.
 Notice that infinitely many other sequences of Misiurewicz points appear
at the band-splitting cascades within the windows of periodic
attractors for values of $\mu > \mu_{\infty }$ \cite{Schuster1}.

\begin{figure*}[t] 
\centering
\epsfig{file=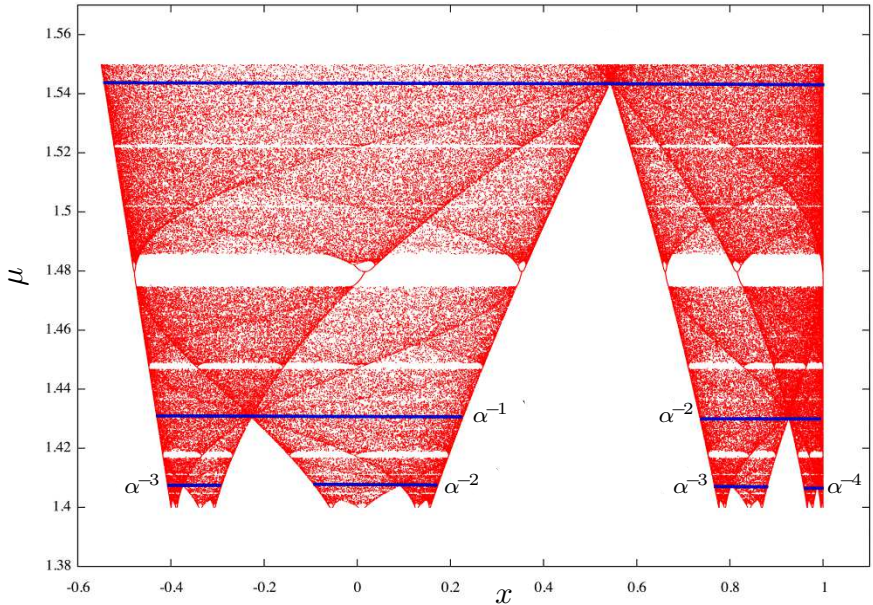, width=.67\textwidth} 
\caption{\scriptsize
  Sector of the main band-splitting cascade for the logistic map
  $f_\mu(x)$ that shows the formation of a Pascal triangle of band widths
  (solid lines) at splitting according to the scaling approximation
  explained in the text, where $\alpha \simeq 2.50291$ is the absolute value of
  Feigenbaum's universal constant.} 
\label{bandSplittingColor}
\end{figure*}

\section{Pascal Triangles and power-law scaling}

The Pascal Triangles and the power-law scaling present in the families
of attractors along the bifurcation cascades of the logistic map are
best appreciated graphically. These can be visualized in Figs. 1 to 4
that we describe below.
 
Fig.\ 1 shows the main period-doubling cascade for the logistic map
$f_\mu(x)$ in the $(x,\mu)$ plane from period two to its accumulation point
(the transition to chaos) and beyond. A few of the supercycle
diameters are shown and their approximate sizes are annotated as
inverse powers of the (absolute value of the) universal constant
$\alpha=-2.50291\ldots$ \ For convenience we denote $|\alpha|$ as $\alpha$
 below and in the figure captions.
 The diameters naturally assemble into well-defined size groups
and the numbers of them in each group can be precisely arranged into a
Pascal Triangle. In fact, the diameter lengths within each group are
not equal, however the differences in lengths within groups diminishes
rapidly as the period $2^n$ increases \cite{Robledo6}. 
A detailed study of the quantitative differences between the values of the 
diameters generated by the logistic map and those obtained from the binomial
approximation that form the Pascal triangle in Fig.\ 1 is given in \cite{Robledo6}.
There are two groups with only
one member, the largest and the shortest diameters, and the numbers
within each group are given by the binomial coefficients.

Fig.\ 2 shows a segment of the main band-splitting cascade of the
logistic map $f_\mu(x)$ in the $(x,\mu)$ plane where we indicate the widths
of these bands at the control parameter values $\hat{\mu}_n$ when they each
split into two new bands. As in the case of the diameters the band
widths diminish in size according to the same inverse powers of $\alpha$
as their numbers $2^n$ increase. They also form groups of nearly equal
sizes with numbers given by the binomial coefficients. Again if it is
assumed that for every value of $n$ the widths of comparable lengths
have equal lengths then these widths can be obtained from the widths
of shortest and longest lengths via a simple scale factor consisting
of an inverse power of $\alpha$.  Under this approximation we observe a
Pascal Triangle across the band-splitting cascade.

Fig.\ 3 shows the period-doubling cascade for the logistic map $f_\mu(x)$
in logarithmic scales in order that the power-law scaling of positions
of trajectories along the supercycle attractors is plainly
observed. These positions are shown as circles. As it can be
appreciated in the figure certain sequences of positions each
belonging different cycles fall on straight lines that share a common
slope. One such alignment corresponds to positions of the principal
diameters $d_{n,0}$ that are formed when $x=0$ is the
other endpoint of the interval. (The positions at $x=0$ do not appear in
the figure as their logarithm is minus infinity). The power-law
scaling is of the form $|x| \sim \alpha^{-n}$.

\begin{figure*}[t] 
\centering
\epsfig{file=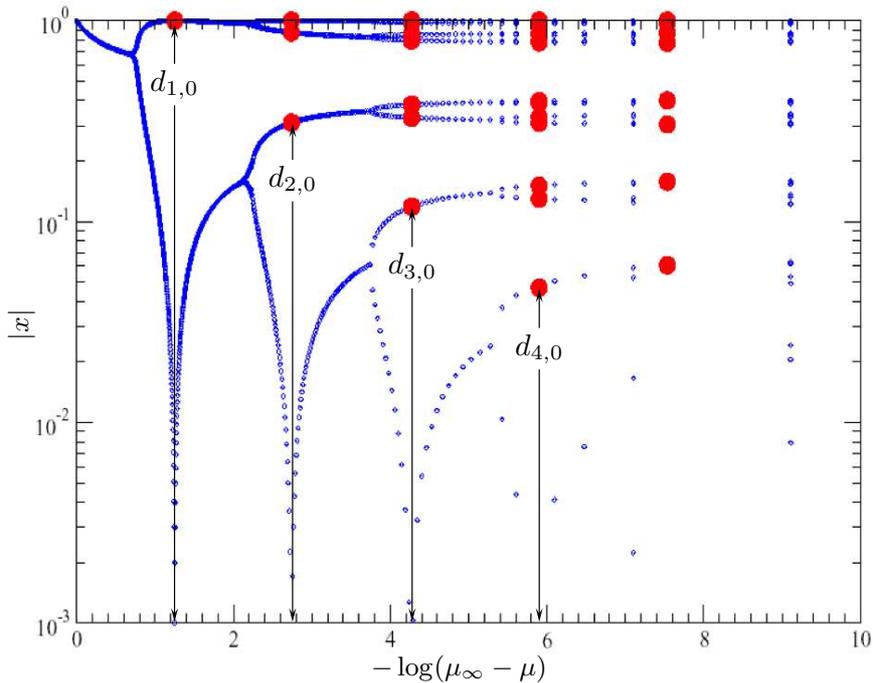, width=.67\textwidth} 
\caption{\scriptsize
  Absolute value of attractor positions for the logistic map
  $f_\mu(x)$ in logarithmic scale as a function of the logarithm of the
  control parameter difference $\mu_\infty - \mu$. The supercycle attractors are
  shown by the arrows and the notation $d_{1,0}, d_{2,0},\ldots$ corresponds to the
  so-called principal diameters, the distances between $x = 0$, the position of the
  maximum of the map, and the nearest cycle position.}
\label{attractorPositions}
\end{figure*}

Fig.\ 4 shows the band-doubling cascade of chaotic attractors in the
logistic map $f_\mu(x)$ in logarithmic scales so that the power-law
scaling of splitting positions $M_n$ is clearly observed. Some of these
positions are shown as circles. As it can be observed in the figure
sequences of Misiurewicz points fall on straight lines that share a
common slope. The power-law scaling is again of the form 
$|x| \sim \alpha^{-n}$.

\begin{figure*}[t] 
\centering
\epsfig{file=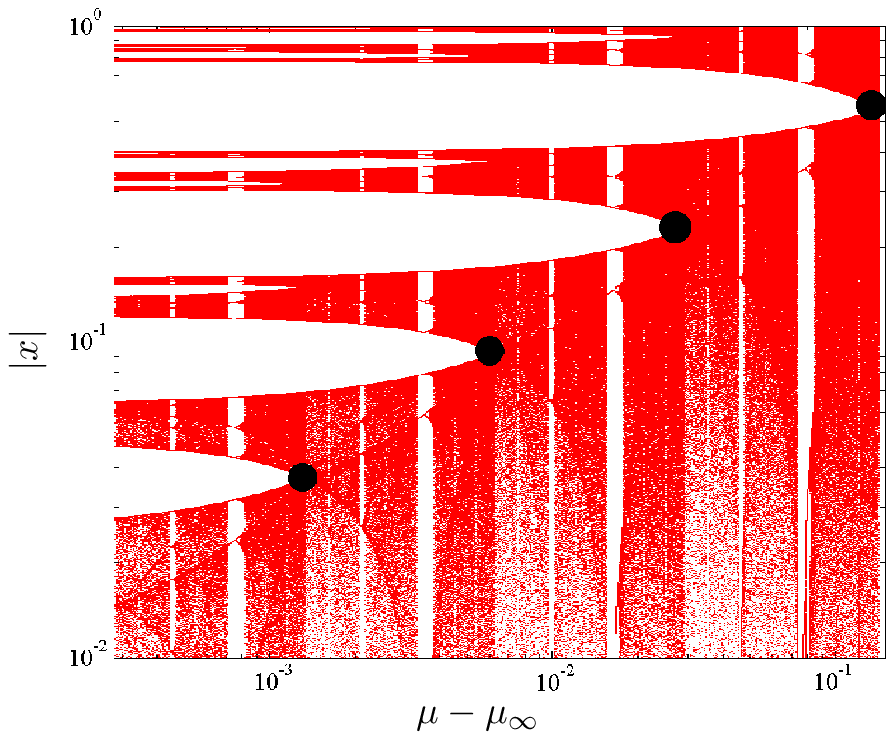, width=.67\textwidth} 
\caption{\scriptsize
  Attractor bands and gaps between them (white horizontal regions) in logarithmic scales,
 $-\log(\mu - \mu_\infty)$ and $\log(|x|)$ in the horizontal and vertical axes, respectively. The band-splitting
 points $M_n$ (circles) follow a straight line indicative of power-law scaling. The vertical white
 strips are periodic attractor windows.}
\label{attractorBands}
\end{figure*}

\section{Gaussian and multi-scale distributions}

As mentioned, the Pascal Triangle is closely associated with a special
case of the Central Limit Theorem, known as the De Moivre-Laplace
theorem that dates back to 1870 \cite{Moivre1}. This theorem establishes that the
limiting form of the binomial distribution, the sum of the binomial
series of $(p+q)^n$ for which the number of successes $s$ falls between $p$
and $q$, $p+q=1$, is the Gaussian distribution. Therefore the occurrences
of approximate Pascal Triangles for the sets of diameters and
bandwidths in the bifurcation cascades implies a Gaussian distribution
for these lengths at their accumulation point, the period-doubling
onset of chaos.  On the other hand, the power-law scaling of
supercycle and band-splitting positions is reflected onto the
period-doubling onset of chaos as a complete organization of
subsequences of positions each following the same power-law
scaling. Iteration time evolution at $\mu_{\infty}$ from $t= 0$ up to
$t\rightarrow\infty$ traces the period-doubling cascade progression from
$\mu = 0$ up to $\mu_{\infty}$. There is a quantitative relationship
between the two developments. Specifically, the trajectory inside the
attractor at $\mu_{\infty}$ with initial condition $x_0 = 0$, the
$2^{\infty}$-supercycle orbit, takes positions $x_t$, such that the
distances between appropriate pairs of them reproduce the diameters
$d_{n,m}$, defined for the supercycle orbits with $\bar{\mu}_n<\mu_{\infty}$
\cite{Robledo1}. See Fig.\ 5, where the absolute value of positions and
logarithmic scales are used to illustrate the equivalence.

The limit distributions of sums of positions at the period-doubling
transition to chaos in unimodal maps have been studied by making use
of the trajectory properties described above \cite{Robledo2},
\cite{Robledo3}. Firstly, the sum of positions as they are visited by a
single trajectory within the attractor was found to have a multifractal structure
imprinted by that of the accumulation point attractor \cite{Robledo2}. It was shown,
analytically and numerically, that the sum of values of positions
display discrete scale invariance fixed jointly by the universal
constant $\alpha$, and by the period doublings contained in the number
of summands. The stationary distribution associated with this sum has
a multifractal support given by the period-doubling accumulation point
attractor \cite{Robledo2}. Secondly, the sum of subsequent positions
generated by an ensemble of uniformly distributed initial conditions
in the entire phase space was recently determined \cite{Robledo3}. It was found that
this sum acquires features of the repellor preimage structure that
dominates the dynamics toward the attractor. The stationary
distribution associated with this ensemble has a hierarchical
structure with multi-scale properties \cite{Robledo3}. See also
Ref.\ \cite{Robledo4}.

\begin{figure*}[t] 
\centering
\epsfig{file=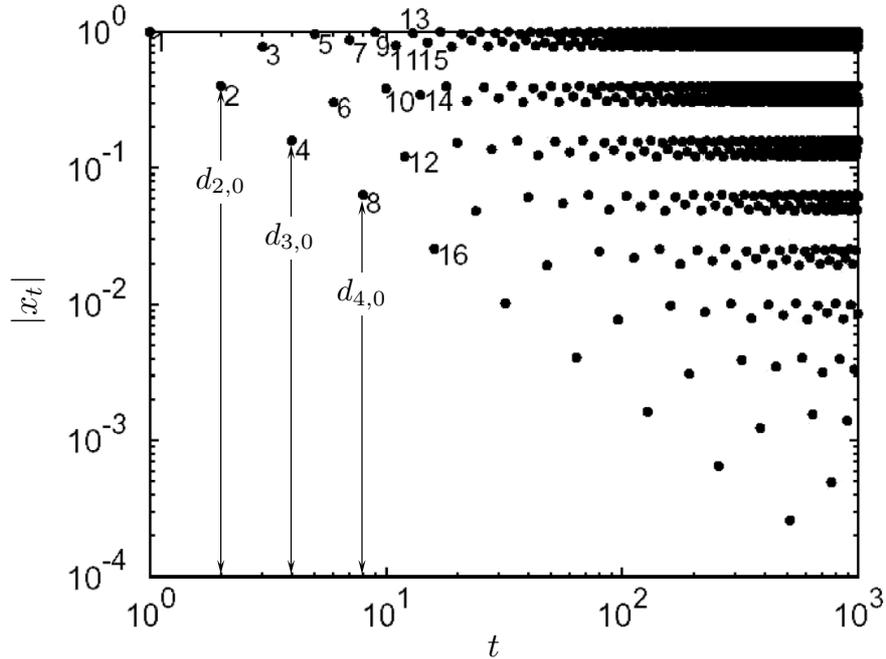, width=.67\textwidth} 
\caption{\scriptsize
  Absolute value of trajectory positions $x_t$, $t = 0,1,\ldots$, for
  the logistic map $f_\mu(x)$ at $\mu_\infty$, with initial condition $x_0 = 0$, in
  logarithmic scale as a function of the logarithm of the time $t$, also
  shown by the numbers close to the points.}
\label{trajectoryPositions}
\end{figure*}

\section{Scale invariant properties and statistical-mechanical structures}

Underlying the power-law scaling of positions are the self-affine
properties that permeate the dynamics of unimodal maps. As we have
seen, these properties manifest visibly along the well-known
bifurcation cascades described here. They are also at the center of
the Renormalization Group (RG) functional composition developed time
ago \cite{Schuster1} to provide a firm theoretical basis to the universal
properties of the accumulation point of the bifurcation cascades. The
RG fixed-point map $f^*(x)$ satisfies the condition
$f^*(f^*(x))= \alpha f^*(x/\alpha)$. Fig.\ 5 offers us a simple visual
opportunity to observe the effect of the RG functional composition and
its fixed-point map property. First we notice that certain position
subsequences appear aligned in the figure.  The most visible is that
of $|x_t|$,  $t=2^n$, $n=0,1,2,3,\ldots$, with a slope equal to $\ln\alpha/\ln 2$,
the main diagonal pattern in the figure. Next to it is the subsequence
$|x_t|$, $t=3\times2^n$, $n=0,1,2,3,\ldots$, with the same slope, and so on. All the
positions $x_t$ can be distributed into subsequences of the form
 $|x_t|$, $t=(2l+1)2^n$, $n=0,1,2,3,\ldots$, $l=0,1,2,3,\ldots$,
 that form a family of lines in the figure
with the fixed slope $-\ln\alpha/\ln 2$ \cite{Robledo1}. And all of the
infinite family of aligned subsequences of positions can be collapsed onto a single
line via rescaling with the use of a ``waiting'' time $t_w=2l+1$, $l=0,1,2,3,\ldots$, a
property known as “aging” in the topic of glassy dynamics \cite{Robledo1},
\cite{Robledo5}.

We observe that the positions in the
logarithmic scales of Fig.\ 5 appear grouped into horizontal bands
separated by gaps of equal widths. The top band contains one half of
the attractor positions as all the odd iteration times appear
there. The second band is made of one quarter of the attractor
positions, those that correspond to iteration times of the form
$t=2+2^n$, $n=0,1,2,3,\ldots$ \  And similarly, the $(k+1)$-th band contains $1/2^n$
of the positions of the attractor, those for iteration times of the
form $t=2^k+2^n$, $n=0,1,2,3,\ldots$, $k=0,1,2,3,\ldots$ \  The RG method
successively transforms the system under study by elimination of
degrees of freedom followed by rescaling aimed at restoring its
original condition. This procedure can be envisaged in Fig.\ 5 by eliminating its
top band, one half of the total attractor positions, and
restoring the original figure by shifting the remaining positions
horizontally by an amount $-\ln 2$ and vertically by an amount
$\ln\alpha$, in line with the slope $-\ln\alpha/\ln2$ of the aligned
subsequences. This graphical procedure is equivalent to functional
composition. Repeated application of this transformation leads
asymptotically to the scaling property of the fixed-point map $f^*(x)$.

The scale invariant features of the accumulation point of the
bifurcation cascades can be related to statistical-mechanical
properties that involve generalized entropy expressions \cite{Robledo1},
\cite{Robledo6}.


\section*{Aknowledgements}
AR is grateful for the hospitality received from the organizers of SPMCS14 in Yichang, China,
and acknowledges support from DGAPA-UNAM-IN103814 and CONACyT-CB-2011-167978 (Mexican Agencies).
CV acknowledges the support provided by IIMAS-UNAM.



\end{document}